\documentclass{vgtc}                          

\ifpdf
  \pdfoutput=1\relax                   
  \usepackage{graphicx}                
  \usepackage{subcaption}
  \usepackage{amsmath}
  \DeclareGraphicsExtensions{.pdf,.png,.jpg,.jpeg} 
\else
  \usepackage{graphicx}                
  \DeclareGraphicsExtensions{.eps}     
\fi%

\graphicspath{{figures/}{pictures/}{images/}{./}} 

\usepackage{microtype}                 
\PassOptionsToPackage{warn}{textcomp}  
\usepackage{textcomp}                  
\usepackage{mathptmx}                  
\usepackage{times}                     
\usepackage{cite}                      
\usepackage{tabu}                      
\usepackage{booktabs}                  
\usepackage[hidelinks]{hyperref}

\onlineid{4133}

\vgtccategory{Research}

\title{A Deep Cybersickness Predictor through Kinematic Data with Encoded Physiological Representation}

\author{Ruichen Li\thanks{e-mail: rli965@connect.hkust-gz.edu.cn}\\ %
       \parbox{2in}{\scriptsize \centering The Hong Kong University of \\ Science and Technology (Guangzhou)}
\and Yuyang Wang\thanks{e-mail: yuyangwang@hkust-gz.edu.cn (Corresponding author)}\\ %
     \parbox{2in}{\scriptsize \centering The Hong Kong University of \\ Science and Technology (Guangzhou)}
\and Handi Yin\thanks{e-mail: hyin335@connect.hkust-gz.edu.cn}\\ %
     \parbox{2in}{\scriptsize \centering The Hong Kong University of \\ Science and Technology (Guangzhou)}
\and Jean-Rémy Chardonnet\thanks{e-mail: jean-remy.chardonnet@ensam.eu}\\ %
     \parbox{2in}{\scriptsize \centering Arts et Metiers Institute of Technology\\ LISPEN, HESAM Universit\'e }
\and Pan Hui\thanks{e-mail: panhui@ust.hk (Corresponding author)}\\ %
     \parbox{2in}{\scriptsize \centering The Hong Kong University of \\ Science and Technology (Guangzhou)}}

\abstract{Users would experience individually different sickness symptoms during or after navigating through an immersive virtual environment, generally known as cybersickness. Previous studies have predicted the severity of cybersickness based on physiological and/or kinematic data. However, compared with kinematic data, physiological data rely heavily on biosensors during the collection, which is inconvenient and limited to a few affordable VR devices. In this work, we proposed a deep neural network to predict cybersickness through kinematic data. We introduced the encoded physiological representation to characterize the individual susceptibility; therefore, the predictor could predict cybersickness only based on a user's kinematic data without counting on biosensors. Fifty-three participants were recruited to attend the user study to collect multimodal data, including kinematic data (navigation speed, head tracking), physiological signals (e.g., electrodermal activity, heart rate), and Simulator Sickness Questionnaire (SSQ). The predictor achieved an accuracy of 97.8\% for cybersickness prediction by involving the pre-computed physiological representation to characterize individual differences, providing much convenience for the current cybersickness measurement.  %
} 

\CCScatlist{
Cybersickness Prediction, VR, Kinematic data, Physiological Representation, Deep Neural Classifiers
}

\begin{document}


\firstsection{Introduction}

\maketitle


Virtual reality (VR) technology has made significant progress in recent years, gaining popularity among many fields such as education, gaming, and healthcare~\cite{van2010survey}. It can create realistic virtual scenes, providing an immersive environment through multimodal feedback for users. However, when users are exposed to VR applications, they may experience cybersickness (CS), which is a common side effect of VR and an unavoidable problem. CS symptoms include headache, eyestrain, stomach awareness, and disorientation~\cite{laviola2000discussion}, and these symptoms can last up to five hours after VR immersion~\cite{regan1994some}. Studies report that more than 60 \% of VR users experience CS, making it a primary impediment to the expansion of the VR market~\cite{lawson2014motion}. To alleviate the influence of CS on VR users, it is vital to develop some strategies to predict it. Predicting the early onset of CS can help us understand the possible causes and potential risk factors so that we can take corresponding strategies for the prevention of severe CS symptoms.

Previous studies have investigated CS prediction utilizing data from integrated sensors in head-mounted displays (HMDs)~\cite{kim2017deep,arcioni2019postural,islam2021cybersickness,chang2021predicting}. These sensors can acquire stereoscopic videos/images and kinematic data, which are then used as input for machine learning algorithms to predict the severity of CS. Kinematic data include head movement, eye movements, and operations~\cite{lencioni2019human}, which have been commonly used as objective cybersickness indicators and have been utilized for predictions~\cite{chang2021predicting,feigl2019sick}. However, kinematic data from different users may demonstrate similar patterns and trajectories in the same VR environment~\cite{wang2021development}, making it challenging to provide deeper insights into individual differences. Similar to kinematic data, stereoscopic videos/images also struggle to accurately reflect individual differences. This is because even with the same visual content, people can have varying interpretations. 

To involve individual differences, CS prediction using physiological data and biometrics has received increasing attention~\cite{silva2019early}. Physiological data are highly valued because of their reliability and objectivity in characterizing individual differences from both physiological and psychological perspectives. Prior studies report that electrodermal activity (EDA) on the forehead has a higher correlation with CS and can be used for the prediction~\cite{yokota2005motion,gavgani2017profiling}. In addition to EDA, other physiological data are widely used for CS prediction, such as heart rate (HR), heart rate variability (HRV), electroencephalography (EEG), and electrogastrogram (EGG)~\cite{islam2020automatic,kim2019deep,dennison2016use}. However, collecting physiological signals requires external biosensors, which presents more challenges in deployment among current affordable consumer-level HMDs such as Oculus Quest 2 and HTC Vive. In addition, the data collection process for physiological signals requires users to limit their interaction experience to prevent noisy information, further complicating the use of external biosensors. To fully unleash the potential of VR, the data collection method should be less intrusive and remain cognitively effortless during the interaction.

To overcome the above-mentioned limitations, we aim to predict the severity of CS only using kinematic data from integrated sensors in HMDs without employing additional biosensors. In addition, we will involve individual differences in the predictive model with EDA-related encoded physiological representation. Kim \textit{et al.} have used stereoscopic videos to estimate EEG-related cognitive state, achieving an accuracy of 90.48\% for CS prediction~\cite{kim2019deep}. However, using stereoscopic videos for prediction has several drawbacks, such as the need for large transmission bandwidth and storage space, privacy concerns, difficulty in analysis, and sophisticated neural network models. Moreover, relying only on video data leads to a narrow perspective, which may lose some key information associated with the CS, leading to worse prediction accuracy. EEG data are recorded from electrodes on the scalp, requiring specialized and expensive equipment, and participants need to prepare by washing their hair and avoiding stimulants before recording. Thus, in this work, we use kinematic data including motion and head-tracking data, which are much convenient for deployment and avoid many privacy concerns. Moreover, instead of using EEG whose collection and processing are more complex and time-consuming, we opt for EDA and propose an EDA-enhanced kinematic model. Specifically, we first encode EDA data into physiological representation. Subsequently, we train an embedding layer to learn physiological embedding from kinematic data, gradually approaching the representation. Finally, our model can interpret individual differences with physiological embedding, without the need for the encoded physiological representation from EDA data. Using the physiological representation to enhance kinematic data provides an in-depth understanding of CS since it captures both inter and intra-individual differences of the CS. In summary, we have made the following three key contributions:

\begin{itemize}
\item Achieve improved performance with only EDA data using temporal features and numerical representation formulated from statistical and topological features.
\item Propose a novel deep learning model to predict CS, which considers kinematic and physiological features using kinematic data only.
\item Insight into the influence of time span and exposure time on CS.
\end{itemize}

\section{Related work}

Cybersickness is a common phenomenon often appearing to VR users during or after an immersive experience. This phenomenon has become an important area of research due to the increasing popularity of immersive technologies and the potential adverse effects on user experience and well-being. In related work, we will review the current understanding of CS, including its causes, measurements, and predicting approaches.

\subsection{Causes}

Cybersickness symptoms are similar to motion sickness, but the underlying causes are different~\cite{stanney1997cybersickness}. Researchers have proposed various theories to explain the mechanisms of CS. One of the most popular theories is the sensory conflict theory, which suggests that CS occurs due to conflicts between visual and vestibular information~\cite{reason1975motion}. The subjective vertical theory proposes that CS occurs due to a conflict between the sensory information from the body and the brain's expectations~\cite{bles1998motion}. To better understand CS, researchers have recommended integrating these two theories into a comprehensive model~\cite{bos2008theory}.

Factors associated with CS can be classified into three categories: hardware, content, and human factors~\cite{chang2020virtual}. Hardware factors such as field of view (FOV) and latency have been widely studied. Adaptive FOV has been proposed as a way to reduce CS, with FOV representing the maximum visual angle of a display device. Kim \textit{et al.} employed a dynamic FOV system that adjusts the FOV using electrophysiological signals. Their results indicate that reducing FOV effectively alleviates discomfort, especially during acceleration and rotational movements~\cite{kim2008application}. Another hardware factor that affects CS is latency. DiZio \textit{et al.} demonstrated that increasing latency increases the severity of VR sickness~\cite{dizio1997circumventing}. Content factors, such as optical flow, graphic realism, and controllability, have also been studied. Optical flow is the VR scene's movement that allows individuals to experience illusory self-motion. Faster locomotion speed in the VR scene is associated with more severe CS~\cite{chardonnet2015visually, liu2012study}. Human factors such as age, ethnicity, and gaming experience significantly correlate with CS levels~\cite{wang2021using}.

\subsection{Measurements}

Several methods have been developed to measure CS, including subjective and objective measures.

\subsubsection{Subjective measurements of CS}

First introduced by Kellogg \textit{et al.}, the Motion Sickness Questionnaire (MSQ) remains a subjective tool for evaluating CS across various dimensions~\cite{kellogg1964motion}. Later, Golding \textit{et al.} proposed a shortened version of the Motion Sickness Susceptibility Questionnaire (MSSQ)~\cite{golding1998motion}. However, the Simulator Sickness Questionnaire (SSQ), developed by Kennedy \textit{et al.}, is widely recognized and universally used as a subjective measurement tool for CS~\cite{kennedy1993simulator}. The SSQ comprises 16 questions divided into three categories (i.e., nausea, oculomotor, disorientation) to assess the severity of each potential manifestation of CS. The scores from these subcategories are combined to determine the total SSQ score, which indicates the overall severity of the sickness. Although the SSQ is a comprehensive and widely used tool, it may lead to an extra workload in answering 16 questions. Therefore, researchers have developed relatively short and quick-to-report questionnaires. For example, the Fast Motion Sickness Scale (FMS)~\cite{keshavarz2011validating} and the Misery Scale (MISC)~\cite{reuten2021feelings} are well-known unidimensional questionnaires. For example, FMS comes with one question asking participants to rate their CS level on a scale ranging from 0 (no sickness) to 20 (severe sickness). The severity of CS experienced by the participant is reported as a score between 0 and 10 by the MISC.

\subsubsection{Objective measurements of CS}

In addition to subjective measurements, researchers have explored objective measurements on top of self-reported evaluation. Two commonly measured indicators are postural sway and electrophysiological changes. Participants are asked to stand on a motion platform to measure postural sway while their VR position is recorded~\cite{villard2008postural,chen2011control}. Multiple studies have demonstrated that these measurements can effectively predict the onset of VR sickness~\cite{dong2010postural,plouzeau2015effect}. Real-time monitoring of participants' psychophysical state changes during the VR experience is also conducted. A significant study by Kim \textit{et al.} examined the physiological indicators of VR sickness, recording various electrophysiological measures before, during, and after the VR experience~\cite{kim2005characteristic}. Among these indicators, electrogastrogram (EGG), eye blink, heart period, and the delta and beta power bands of electroencephalogram (EEG) exhibit CS-specific responses. Additionally, phasic skin conductance at the forehead, as reported by Golding and John ~\cite{golding1992phasic}, has also shown a strong correlation with motion sickness onset. In a subsequent study, Gavgani \textit{et al.}~\cite{gavgani2017profiling} confirmed that skin conductance on the forehead is a reliable measure of nausea during immersion. Other studies have reported that CS can cause a 4$^{\circ}$C increase in body temperature and a 20-50ms increase in reaction time~\cite{nalivaiko2015cybersickness}. Furthermore, researchers have examined blood pressure, fMRI, theta/total parameter at the frontal and parietal lobes, and hormonal levels as potential measures of VR sickness.

\subsection{Predicting Approaches}

Previous studies have explored using Machine Learning (ML) and Deep Learning (DL) models to predict CS based on individual factors, stereoscopic 3D videos, kinematic data, and physiological signals. According to the data source, these approaches can be divided into two categories.

\subsubsection{CS Predicting with integrated sensors in HMDs}

Integrating various sensors within VR HMDs allows for collecting multimodal data, including videos, images, and user kinematic data. These sensors track the user's head movement, operation, and body position, which are then processed by the system to create a seamless VR experience. 

Regarding videos, Padmanaban~\textit{et al.}~\cite{padmanaban2018towards} created a dataset of 19 stereoscopic videos and corresponding sickness levels to predict simulator sickness. They considered FOV, velocity, and stimulus depth as features and evaluated them with a decision tree model. The prediction performance was later improved with a 3D CNN model that used saliency, disparity, and optical flow features~\cite{lee2019motion}. Du \textit{et al.} built a CNN model that automatically learnt and integrated multi-level stereoscopic visual sickness descriptors for simulator sickness prediction~\cite{du2021learning}. 

In addition to utilizing video data for prediction, researchers have also leveraged user kinematic data, including eye-tracking~\cite{guo2011could}, motion~\cite{hadadi2022prediction,feigl2019sick}, head-tracking~\cite{jin2018automatic} and postural data~\cite{arcioni2019postural,risi2019effects,widdowson2021assessing}. Gazing activity has been found to help reduce CS severity by Chang \textit{et al.}\cite{chang2021predicting}. Lopes \textit{et al.}~\cite{lopes2020eye} found that the CS level can be affected by both the position of the pupil and the type of blink. Wang \textit{et al.} utilized normal-state postural data collected from participants to train a long short-term memory (LSTM) model, which was then used to predict CS levels~\cite{wang2019vr}. Islam \textit{et al.} proposed a novel approach for assessing CS severity using a multimodal deep fusion network that integrates data from various modalities, including eye-tracking, head-tracking, and stereo-image data~\cite{islam2021cybersickness}.

\subsubsection{CS Predicting with external biosensors}

Previous research has suggested various approaches to predict CS by utilizing physiological signals, which have been found to be significantly correlated with CS. Garcia \textit{et al.} conducted an experiment with 66 participants, collecting electrocardiographic, electrooculographic, respiratory, and skin conductivity data to classify cybersickness using KNN and SVM classifiers~\cite{garcia2019development}. The accuracy achieved by the Binary and Ternary classifiers were 82\% and 56\%, respectively. Kim \textit{et al.} achieved an accuracy of 89.16\% in predicting CS by using EEG data~\cite{kim2019deep}. Their study involved collecting EEG data from 200 participants who were immersed in 44 virtual reality (VR) simulations using an 8-channel EEG. The authors estimated the cognitive state and its association with CS levels using CNN and LSTM models based on brain signals. Jeong \textit{et al.} proposed a data processing technique for DNN models and achieved an impressive accuracy rate of 98.02\% using EEG data. Additionally, they observed a distinct pattern of CS occurrence during VR simulations.


Besides EEG, electrodermal activity (EDA) is a widely used measure of emotional arousal. It includes Skin Conductance Level (SCL) and Skin Conductance Response (SCR)\cite{imotions2017galvanic}. The SCL changes slightly on a time scale of tens of seconds to minutes and can vary between individuals. The SCR, on the other hand, is the phasic component that shows rapid changes and is sensitive to emotionally arousing events, such as event-related SCRs (ER-SCRs). These variations in the phasic component are visible as EDA bursts or peaks, which can occur within 1-5 seconds after the onset of emotional stimuli. Islam \textit{et al.} achieved an 87.38\% accuracy in predicting CS by analyzing heart rate (HR) and EDA data from 22 participants using deep neural networks\cite{islam2020automatic}. They also proposed a multimodal fusion network involving eye tracking, HR, and EDA to forecast CS onset 60 seconds in advance with a root-mean-square error of 0.49~\cite{islam2022towards}.


\section{User Study}


\begin{figure}[h]
\centering
\includegraphics[width=0.6\columnwidth]{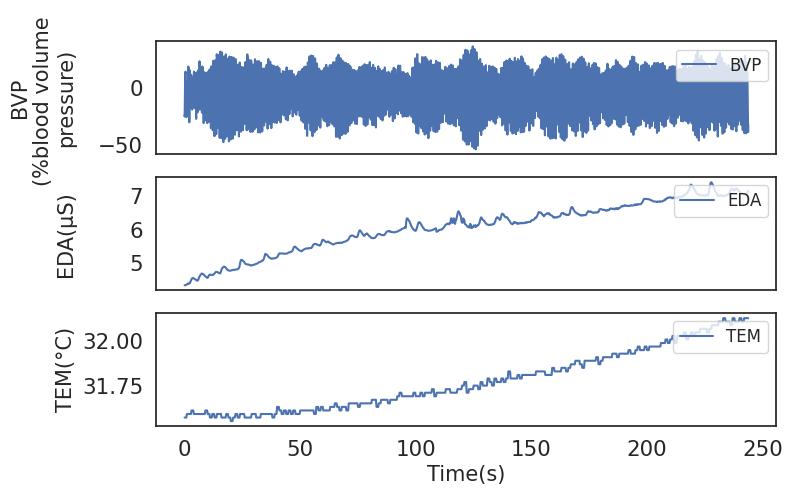}
\caption{Physiological data of one sample.}
\label{fig:physiological} %
\end{figure}

\begin{figure}[h]
\centering
\includegraphics[width=0.6\columnwidth]{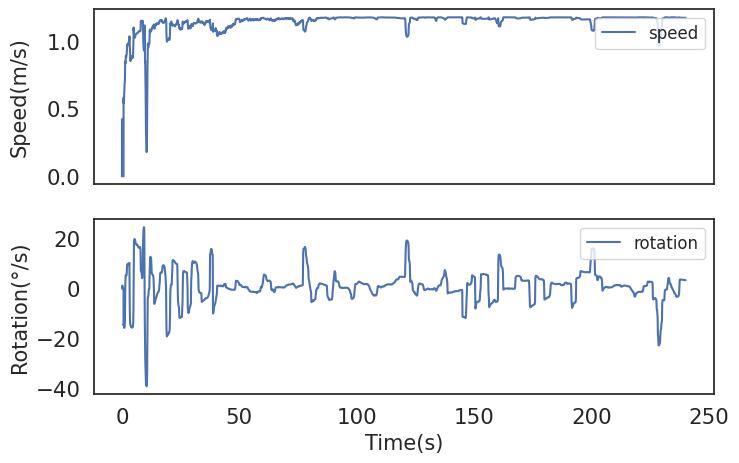}
\caption{Motion data of one sample.}
\label{fig:motion} %
\end{figure}


To train a high-quality model to make CS predictions based on physiological and kinematic data, we conducted a user study to collect the data as ground truth. The task was performed in the HTC Vive Pro Eye. Participant information and task design are presented in \autoref{Participants and Task Design}, and we further introduce the data collection process shown in \autoref{Data Collection Procedure}.

\subsection{Participants and Task Design} \label{Participants and Task Design}

In this study, 53 participants ($M_{age}=26.3$, $SD_{age}=3.3$, females: 26) completed a navigation task in an immersive environment. Most participants were young individuals as VR applications are more popular among this group. Each participant completed the task three times over three days, resulting in 159 samples. However, two samples were discarded due to device defects, leaving 157 samples for data analysis.



Upon arrival, participants signed a consent form indicating potential risks and their agreement for the experiment. They completed a questionnaire assessing their health conditions and experience with gaming and VR devices. All participants confirmed their eligibility, reporting no pre-existing medical issues that would affect their performance. They were compensated with various gifts upon completion of the study.

\begin{figure}[h]
\centering
\includegraphics[width=0.6\columnwidth]{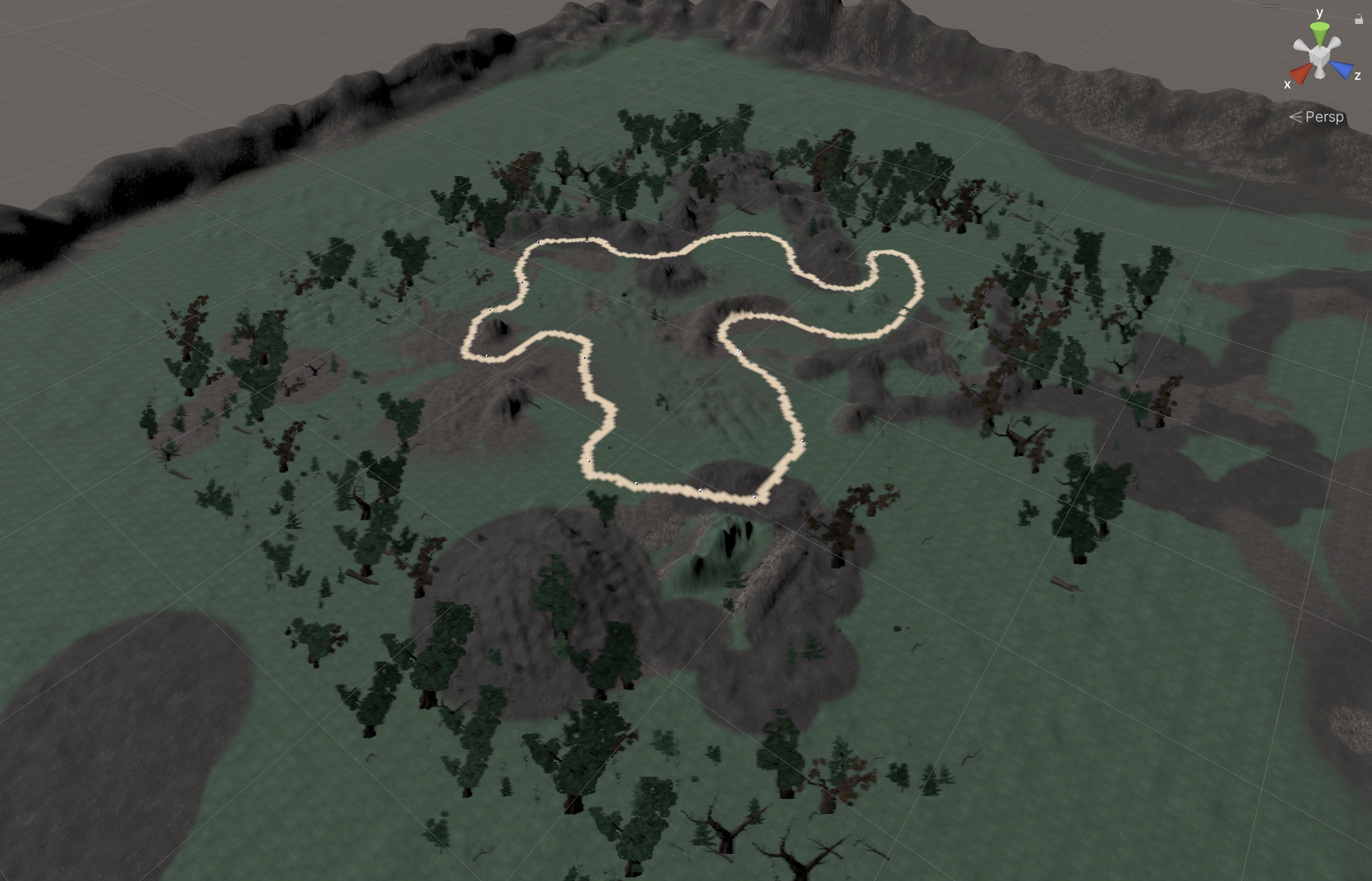}
\caption{Virtual scenario in which the participants navigate along the highlighted path.}
\label{fig:ve} %
\end{figure}

Participants received clear instructions on the navigation task and how to use HTC Vive Pro hand controllers. They had to navigate a virtual forest along a winding gravel path. The VR immersion lasted four minutes to induce moderate cybersickness without posing health risks. \autoref{fig:ve} shows a visual representation of the path.

\subsection{Data Collection Procedure} \label{Data Collection Procedure}

\begin{figure}[h]
\centering
\includegraphics[width=0.9\columnwidth]{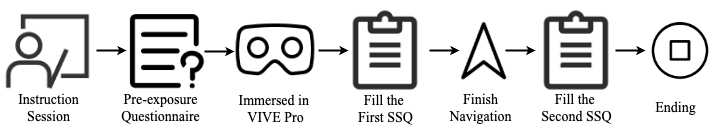}
\caption{Experiment Procedure for the data collection}
\label{fig:process} %
\end{figure}

The general experimental procedure is presented in \autoref{fig:process}.

\begin{enumerate}
   \item Participants received instructions on using HTC Vive Pro hand controllers and were informed they could terminate the experiment if experiencing CS symptoms.
  \item Participants wore an HTC Vive Pro headset and an Empatica E4 wristband on one arm. Real-time physiological data (EDA, BVP, TEM) were collected using the Empatica E4 wristband~\footnote{https://www.empatica.com/research/e4/} with sensors transmitting data via Bluetooth to a processing computer with sampling frequencies: EDA (4Hz), BVP (64Hz), TEM (4Hz). One physiological sample is shown in \autoref{fig:physiological}.
  \item Participants engaged in the virtual environment (\autoref{fig:ve}), directing movements using the HTC Vive Pro hand controller touchpad. Physiological and kinematic data, including head-tracking and motion data, were collected synchronously during four-minute navigation. After completion, participants removed the head-mounted display. One sample of motion data is illustrated in \autoref{fig:motion}.
  \item The Simulator Sickness Questionnaire (SSQ) assessed CS severity, measuring nausea, oculomotor, and disorientation symptoms through 16 questions. Participants completed pre- and post-experiment SSQ to report CS degree. SSQ score calculated by subtracting post-exposure from pre-exposure score, given as, 

  \begin{equation}
      SSQ = SSQ_{post} - SSQ_{pre}
  \end{equation}

\end{enumerate}

We gathered three distinct sets of data, each of which shared the same starting and ending times: head-tracking, motion, and physiological data. The data collected from the integrated sensors of the HMD and Empatica E4 wristband are summarized in \autoref{tab:raw data}.

\begin{table}[htbp]
\centering
\caption{List of temporal data collected during the navigation task}
\label{tab:raw data}
\renewcommand\arraystretch{1.2}
\scalebox{0.7}{
\begin{tabular}{cl}
\toprule
\textbf{Data Type}                                                           & \multicolumn{1}{c}{\textbf{Data}}                                                                                    \\ \hline
\textbf{\begin{tabular}[c]{@{}c@{}}Head-Tracking Data\\ (90Hz)\end{tabular}} & \begin{tabular}[c]{@{}l@{}}Head Position (i.e., x, y and z)\\ Head Rotation (i.e., x, y and z)\end{tabular}          \\ \hline
\textbf{\begin{tabular}[c]{@{}c@{}}Motion Data\\ (90Hz)\end{tabular}}        & \begin{tabular}[c]{@{}l@{}}Speed \\ Rotation\end{tabular}                                                            \\ \hline
\textbf{Physiological Data}                                                      & \begin{tabular}[c]{@{}l@{}}Electrodermal Activity (4Hz)\\ Temperature (4Hz)\\ Blood Volume Pulse (64Hz)\end{tabular} \\ \bottomrule
\end{tabular}
}
\end{table}

\section{Data Processing}

It is necessary to pre-process the raw data from the experiment to construct the input and output for building deep learning models. First, we explain how to establish the ground truth in \autoref{Ground Truth Construction}, then define the CS prediction task in \autoref{Problem Definition}. Next, we show the data processing details in \autoref{Data Processing}. By carefully handling the data, we can ensure that the deep learning model is trained on high-quality inputs and outputs, which will improve its performance and accuracy. 

\subsection{Ground Truth Construction} \label{Ground Truth Construction}

SSQ covers a wide range of symptoms that are commonly associated with CS and it is a standardized tool that has been used as ground truth in many studies~\cite{hadadi2022prediction, jin2018automatic}. So our work used SSQ scores to construct the ground truth. Stanney \textit{et al.} stated that cybersickness could be classified using the following criteria: SSQ scores below 5 indicate negligible symptoms, while scores above 20 suggest a bad intervention~\cite{stanney1997cybersickness}. However, a recent meta-analysis challenges using scores above 20 as an indicator of a bad intervention, as it found that approximately one third of users discontinued with a weighted SSQ score of 40 or higher~\cite{caserman2021cybersickness}. Therefore, we employ a revised classification rule for cybersickness level, which includes four categories: negligible ($0\leq SSQ \leq 5$), low ($5\textless SSQ \leq 20$), moderate ($20\textless SSQ \leq 40$), and high ($SSQ\textgreater 40$). This updated classification standard provides a more accurate reflection of cybersickness level. Also, the samples were uniformly distributed among such classification,  with 30, 41, 53, and 33 samples for each class. The data was labeled as follows:

\begin{equation}\label{eq:sick_group}
CS=\left\{\begin{array}{lcl}
Negligible & , { if } & 0\leq SSQ \leq 5\\
Low & , { if } & 5\textless SSQ \leq 20\\
Moderate & , { if } & 20\textless SSQ \leq 40\\
High & , { if } &  SSQ\textgreater 40 \\
\end{array}\right.
\end{equation}

\subsection{Problem Definition} \label{Problem Definition}

\begin{figure}[h]
\centering
\includegraphics[width=0.9\columnwidth]{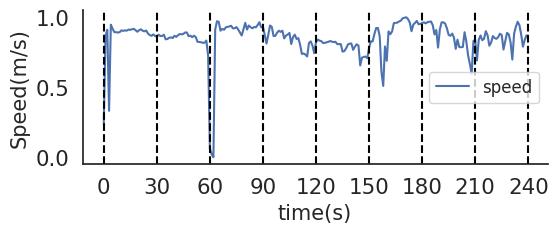}
\caption{The original four minutes are divided into eight 30s segments.}
\label{fig:window_split} %
\end{figure}

Previous studies indicated that the exposure time does not affect the severity of the VR experience. For example, Melo \textit{et al.}~\cite{melo2018presence} found no significant difference between shorter and longer (5 or 7 min) experiences concerning participants' sense of presence and CS. Katharina \textit{et al.}\cite{petri2020effects} also reported no CS differences between 10-minute and 20-minute virtual immersion for karatekas aged 18-60. Thus, if a participant experiences sickness after the four-minute VR experience, we assume the CS level ($CS$) remains constant for all observed data within that time. We will investigate the exposure time-CS relationship further in~\autoref{Performance on different exposure time}. The entire four-minute VR experience is labeled as $CS_{240}$.

\begin{equation}
[D^1, D^2, D^3, D^4, ..., D^{239}, D^{240}] \rightarrow CS \nonumber \end{equation}

$D^t$ represents observed data at time $t$, comprising 4 data points from EDA and Temperature (4Hz), 64 data points from BVP (64Hz), and 90 data points from head-tracking and motion (90Hz). The original data was segmented based on time span $T_s$. For instance, using $T_s=30$s, the four-minute data was divided into eight segments (see \autoref{fig:window_split}). Each time segment was labeled with CS level $CS_{240}$.

\begin{equation}
	\begin{split}
	[D^1, D^2, ..., D^{29}, D^{30}] \rightarrow CS\\
[D^{31}, D^{32}, ..., D^{59}, D^{60}] \rightarrow CS\\
[D^{61}, D^{62}, ..., D^{89}, D^{90}] \rightarrow CS \\
[D^{91}, D^{92}, ..., D^{119}, D^{120}] \rightarrow CS\\
[D^{121}, D^{122}, ..., D^{149}, D^{150}] \rightarrow CS \\
[D^{151}, D^{152}, ..., D^{179}, D^{180}] \rightarrow CS\\
[D^{181}, D^{182}, ..., D^{209}, D^{210}] \rightarrow CS\\
[D^{211}, D^{212}, ..., D^{239}, D^{240}] \rightarrow CS   \nonumber
	\end{split}
\end{equation}

The CS level prediction task can be defined as giving the observed data (e.g., physiological, head-tracking, and motion data) in the time segment with a time span of $T_s$ from a person, determining the user's CS level. 

\subsection{Data Processing} \label{Data Processing}

\begin{table}[!h]
\centering
\caption{An independent t-test of the physiological data between higher CS and lower CS group (df = 157), p-value $<$ 0.05 represents significant properties}
\label{tab:ttest-phy}
\renewcommand\arraystretch{1.2}
\scalebox{0.7}{
\begin{tabular}{lllll}
\toprule
\multicolumn{1}{c}{}  & \multicolumn{1}{c}{\textbf{{Mean}}}    & \multicolumn{1}{c}{\textbf{{SD}}}    & \multicolumn{1}{c}{\textbf{{t-value}}} & \multicolumn{1}{c}{\textbf{{p-value}}} \\ \hline
\multicolumn{1}{c}{\begin{tabular}[c]{@{}c@{}}\textbf{Low-CS BVP}\\ \textbf{High-CS BVP}\end{tabular}} & \multicolumn{1}{c}{\begin{tabular}[c]{@{}c@{}}-0.013569\\ -0.075292\end{tabular}} & \multicolumn{1}{c}{\begin{tabular}[c]{@{}c@{}}0.281273\\ 0.270550\end{tabular}} & \multicolumn{1}{c}{-1.4065} & \multicolumn{1}{c}{0.1616}  \\ \hline
\begin{tabular}[c]{@{}l@{}}\textbf{Low-CS EDA}\\ \textbf{High-CS EDA}\end{tabular}                     & \begin{tabular}[c]{@{}l@{}}2.753912\\ 4.373491\end{tabular}                       & \begin{tabular}[c]{@{}l@{}}3.253526\\ 4.158797\end{tabular}                     & 2.6919                      & \textbf{0.0079}                      \\ \hline
\begin{tabular}[c]{@{}l@{}}\textbf{Low-CS TEM}\\ \textbf{High-CS TEM}\end{tabular}                     & \begin{tabular}[c]{@{}l@{}}34.429848\\ 34.587766\end{tabular}                     & \begin{tabular}[c]{@{}l@{}}1.298672\\ 1.467705\end{tabular}                     & 0.7111                      & 0.4781                      \\ \bottomrule
\end{tabular}
}
\end{table}

\begin{table}[!h]
\centering
\caption{Feature set of one sample}
\label{tab:featureset}
\renewcommand\arraystretch{1.2}
\scalebox{0.7}{
\begin{tabular}{cll}
\toprule
Feature Type   & \multicolumn{1}{l}{\begin{tabular}[c]{@{}l@{}}Data Type\\ (Dimension)\end{tabular}} & \multicolumn{1}{l}{Features}     \\ \hline
Temporal & \begin{tabular}[c]{@{}l@{}}Head-tracking\\ (12)\end{tabular}                 & \begin{tabular}[c]{@{}l@{}}Head Position \\ First difference of Head Position \\ Head Rotation \\ First difference of Head Rotation \\(i.e., x, y and z)\end{tabular}                             \\ \cline{2-3} 
                             & \begin{tabular}[c]{@{}l@{}}Motion\\ (4)\end{tabular}                        & \begin{tabular}[c]{@{}l@{}}Speed \\ First difference of Speed\\ Rotation\\ First difference of Rotation\end{tabular}                                                                                                                                 \\ \cline{2-3} 
                             & \begin{tabular}[c]{@{}l@{}}EDA\\ (15)\end{tabular}        & \begin{tabular}[c]{@{}l@{}}EDA, SCR, SCL\\ $EDA_{min}$, $EDA_{max}$, $EDA_{mean}$, $EDA_{std}$\\ $SCR_{min}$, $SCR_{max}$, $SCR_{mean}$, $SCR_{std}$\\ $SCL_{min}$, $SCL_{max}$, $SCL_{mean}$, $SCL_{std}$\end{tabular}                                                                                           \\ \hline
Numerical                    & \begin{tabular}[c]{@{}l@{}}EDA\\ (38)\end{tabular}       & \begin{tabular}[c]{@{}l@{}}mean\_scl, std\_scl, \\std\_scr, corr, \\ num\_responses, ...\end{tabular}                                                               \\ \bottomrule
\end{tabular}
}
\end{table}

The rendering frequency of Unity3D fluctuates around 90Hz due to computer performance, posing a challenge for data collection at a constant rate. Hence, all head-tracking and motion data were resampled to 90Hz during preprocessing. Subsequently, the following steps were carried out:

\begin{itemize}
    \item \textbf{Correlation Analysis.}   
    We combined the first and last two CS level categories, resulting in the higher CS group ($SSQ > 20$) and lower CS group ($0 \leq SSQ \leq 20$). We conducted independent sample t-tests on the physiological data of these two groups. The results showed a significant difference in EDA signals, as presented in \autoref{tab:ttest-phy}. However, there was no significant difference in BVP and TEM data between the higher and lower CS groups. Therefore, we excluded BVP and TEM signals from further data processing.
    \item \textbf{Data Downsampling.} 
        To account for varying sampling rates, all data was downsampled to 1Hz. The mean value of the samples was used to retain information during downsampling. For example, head-tracking and motion data were aggregated using the mean value over 90 data points per second, while EDA signals were aggregated using the mean value over 4 data points.
    \item \textbf{Data Segmentation.} 
        In order to build our prediction task, we divided the original four-minute data into several windows according to the time span $T_s$. Since $T_s$ is temporarily set to 30s, we got a total of $8*157=1256$ samples, as presented in \autoref{fig:window_split}. 
    \item \textbf{Feature Selection.} 
    We extracted features from head-tracking, motion, and EDA data. The features extracted from one sample are listed in \autoref{tab:featureset}. 
    \begin{itemize} 
        \item For speed and rotation data, we computed the first difference between consecutive frames to calculate the difference between adjacent seconds. For head-tracking vectors (x, y, and z), we computed the first difference of consecutive vectors.
        \item For temporal features of EDA data, courtney \textit{et al.} suggested using a 3-second window to capture changes in EDA data~\cite{courtney2010better}. So we employed a trailing moving average with a 3-second window to smooth the EDA data and account for noise. Properties of physiological signals within the previous 3-second window, including mean, std, maximum, and minimum, were considered indicators of physiological changes. $D_{max}$, $D_{min}$, $D_{mean}$, and $D_{std}$ captured abrupt fluctuations in physiological signals over a 3-second period. As the EDA signals' phasic component (SCR) is associated with arousing stimulus events and the SCL differs significantly across different individuals, we should link the CS level to them. Using Neurokit2~\cite{Makowski2021neurokit}, a Python toolbox designed to process physiological signals, we could extract the SCR and SCL from the original EDA. Then we did the same processing as the original EDA data on the extracted two components. 
        \item For EDA data, in addition to temporal features, we also analyzed each time segment and extracted numerical representation  formulated from statistical and topological features. The meanings of 38-dimension numerical representation are shown in our supplemental material~\footnote{Available on OSF at \href{https://osf.io/9tucz/?view_only=fcf2188c009446cca87aeac085b411d9}{osf.io}, released under a CC BY 4.0 license. They include (1) Excel containing the meanings of 38-dimension numerical features, (2) an example of fine-tuning results, (3) datasets, and (4) source code.}.
    \end{itemize}
    \item \textbf{Data Normalization.} 
        To mitigate the impact of individual differences that could potentially hinder the model's training due to varying signal magnitudes, we opted to normalize all temporal features using min-max normalization, thereby scaling the values to fall within the range of $(0,1)$.
\end{itemize}

\section{Model Architecture}




\begin{figure}[!h]
    \centering
    \includegraphics[scale=0.35]{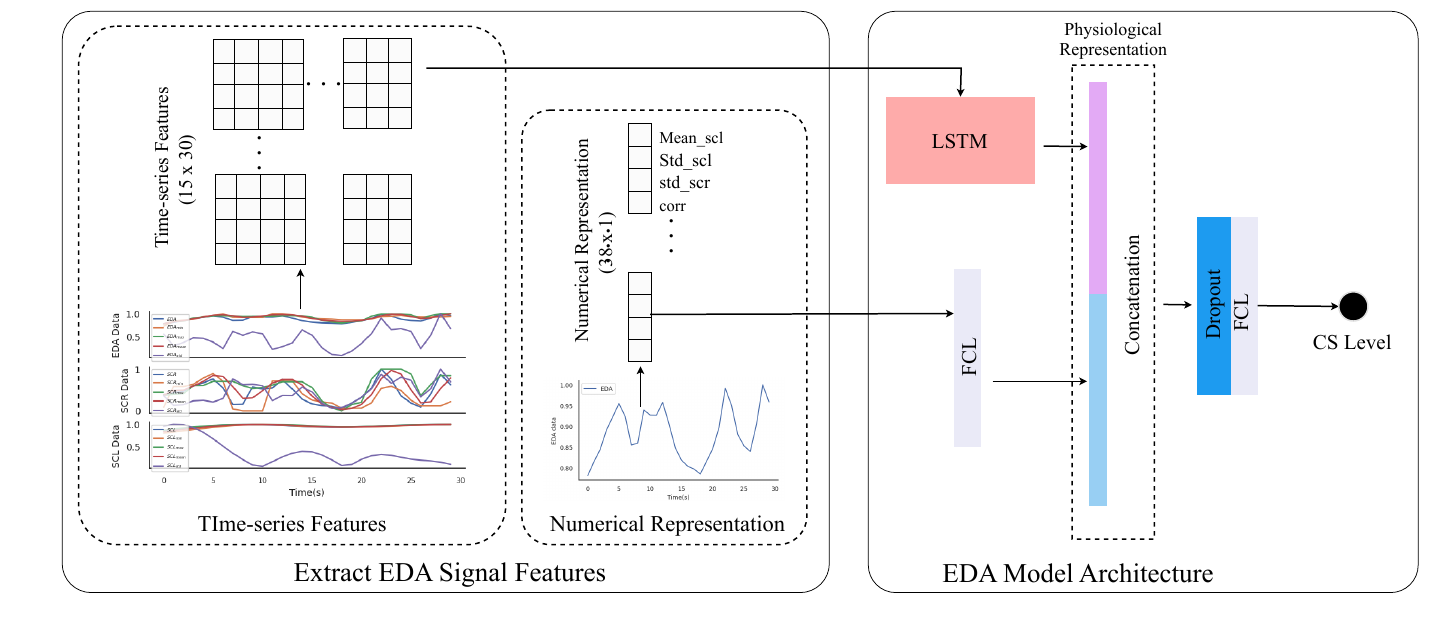}
    \caption{EDA Model. The model aims to encode EDA data as a low-dimensional physiological representation.}
    \label{fig:EDA_model}
\end{figure}

\begin{figure}[!h]
    \centering
    \includegraphics[scale=0.42]{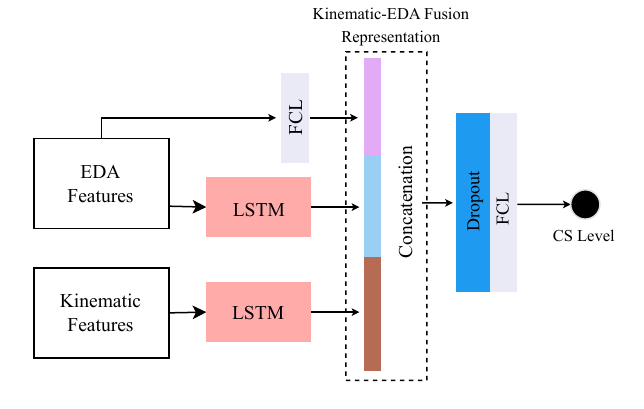}
    \caption{Kinematic-EDA Fusion model. The model learns features from physiological and kinematic data.}
    \label{fig:Fusion_model}
\end{figure}

\begin{figure}[!h]

    \centering
    \includegraphics[scale=0.42]{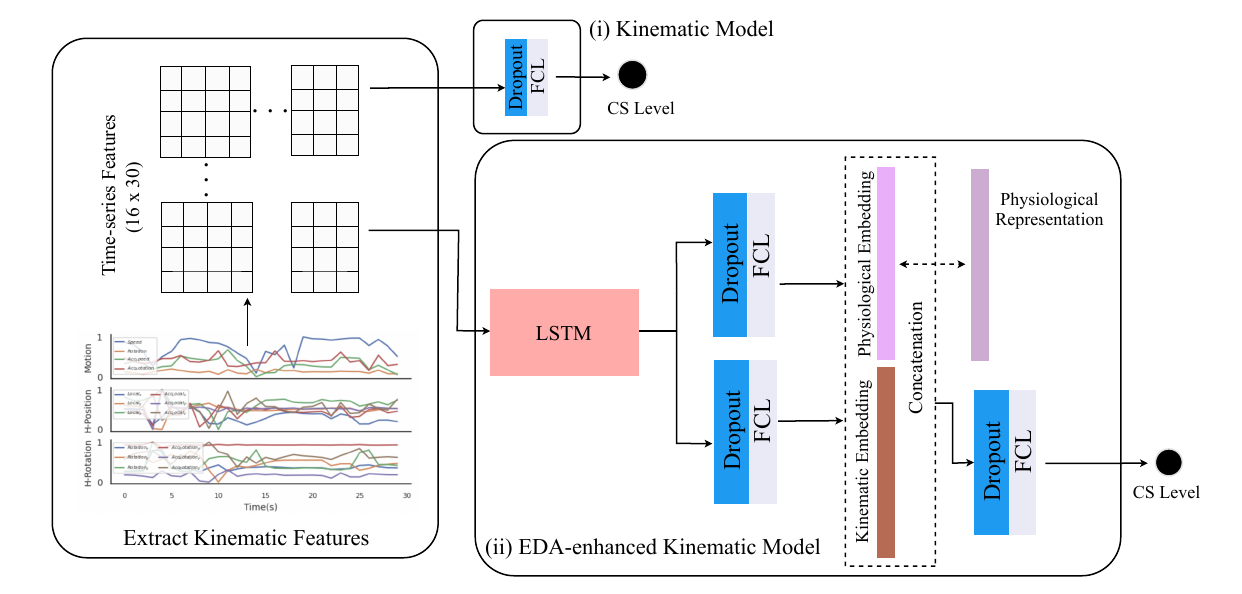}
    \caption{(i) Kinematic Model. The model learns features from kinematic data, including head-tracking and motion data. (ii) EDA-enhanced Kinematic Model.The model learns the physiological and kinematic embedding from Kinematic data.}
    \label{fig:stage1}

\end{figure}

After preparing input and output data, we constructed and evaluated multiple deep learning models for different input modalities, identifying relevant inputs for accurate predictions. Analyzing model performance provided valuable insights into the influence of input features on outcomes, leading to the development of precise and efficient deep learning models for diverse applications. This section describes the environment setup (see \autoref{Setup}) before exploring specific model architectures for different input combinations in subsequent sections.

\subsection{Setup} \label{Setup}

Our deep learning models were trained and evaluated using TensorFlow. Execution took place on a system running Ubuntu 18.04, equipped with two NVIDIA GeForce RTX 3090 Ti GPUs (each with 24GB memory). The dataset underwent 5-fold cross-validation, with 4 groups for training and the remaining group for testing. Mean accuracy served as the primary evaluation metric. To optimize parameters and improve performance, we employed the Optuna~\footnote{https://optuna.org} Python library, which efficiently explores and exploits the parameter space. This significantly enhanced our hyperparameter tuning process and resulted in improved model performance. Additionally, we utilized the Adam optimizer with an epoch of 100 and a batch size of 64.

\subsection{EDA Model} \label{EDA Model}

The EDA model encodes EDA data as a low-dimensional physiological representation. Fig. \ref{fig:EDA_model} shows the model's schematic architecture. The time span $T_s$ is 30s, resulting in a temporal EDA feature size of $(15 * 30)$ and a numerical representation size of $(1 * 38)$. An LSTM layer with ($LSTM\_size$) units learns the temporal representation, while a fully-connected layer with ($Dense\_size_1$) neurons and activation function ($Acti$) learns the numerical representation. These representations are combined using a concatenate layer, forming the physiological representation used as the ground truth in the EDA-enhanced kinematic model discussed later. To prevent overfitting, a dropout layer with a dropout rate of $Rate$ is applied, followed by a fully-connected Dense layer with ($Dense\_size_2$) neurons and activation function ($Acti$). 

\subsection{Kinematic Model} \label{kinematic Model} 

The kinematic model learns features from kinematic data, including head-tracking and motion data. The detailed architecture is depicted in Fig. \ref{fig:stage1}(i). The time span $T_s$ is set to 30s, so the temporal kinematic feature is $(16 * 30)$. We use an LSTM layer with ($LSTM\_size$) units to learn the temporal representation. The output is then passed to a dropout layer with a specified dropout rate $Rate$ to reduce overfitting. This is followed by a fully-connected Dense layer with a defined number of neurons ($Dense\_size$) and an activation function ($Acti$). 

\subsection{Kinematic-EDA Fusion Model} \label{kinematic-EDA Fusion Model}

The kinematic-EDA fusion model combines physiological and kinematic data. See Fig. \ref{fig:Fusion_model} for the detailed architecture. With a time span of 30s, the temporal input EDA feature size is (30 * 15), the temporal input kinematic feature size is (30 * 16), and the numerical input EDA feature size is (1 * 38). Two LSTM layers, with ($LSTM\_size_1$) units for temporal EDA representation and ($LSTM\_size_2$) units for kinematic representation, are employed. A fully-connected Dense layer with ($Dense\_size_1$) neurons and activation function ($Acti_1$) encodes numerical EDA features. The outputs of these three layers are concatenated and passed through a dropout layer with a specified dropout rate. They are then fed into a fully-connected Dense layer with ($Dense\_size_2$) neurons and activation function ($Acti_2$). 

\subsection{Proposed EDA-enhanced Kinematic Model} \label{EDA-enhanced Kinematic Model}

The proposed EDA-enhanced kinematic model aimed to learn the physiological embedding from Kinematic data with encoded physiological representation learned from the EDA model mentioned in \autoref{EDA Model}. The overall architecture is illustrated in Fig. \ref{fig:stage1}(ii). Here we use an LSTM network with ($LSTM\_size$) units as a sequential network to process the kinematic features. Then the dropout layer with the $Rate_1$ and fully-connected Dense layer with ($Dense\_size_1$) and activation function ($Acti_1$) output physiological embedding. Next, the dropout layer with the $Rate_2$ and fully-connected Dense layer with ($Dense\_size_2$) and activation function ($Acti_2$) are used to extract kinematic representation. Then the physiological embedding is concatenated with the kinematic representation to produce the final representation vector. The output is forwarded into the dropout layer with the $Rate_3$ and then passed to a fully connected Dense layer with ($Dense\_size_3$) neurons and activation function ($Acti_3$). The model uses the Adam optimizer with the $Lr$. Note that in order to make the learned embedding  closer to the encoded physiological representation, we design a custom loss function, which is formed by two terms: the prediction loss and the regression loss,

$$ L = L_{pre} + \beta * L_{reg} $$

Where $L_{pre}$ represents the standard cross-entropy loss between ground truth and the output unit of the last FCL, $L_{reg}$ denotes the mean squared error (MSE) or mean absolute error (MAE) between the physiological features and the physiological representation learned by EDA model, and a constant $\beta$ is used to adjust the balance between the two terms. These two parameters are fine-tuned with Optuna.

\section{Experiments and Results}

\begin{table*}[!h]
\centering
\caption{Results of models with different time spans before downsampling}
\label{tab:best_result before downsampling}
\renewcommand\arraystretch{1.2}
\scalebox{1}{
\resizebox{\linewidth}{!}{

{\begin{tabular}{lccccc}
\toprule
 Time span $T_s$  & \begin{tabular}[c]{@{}c@{}}10\\ $mean\pm std$\end{tabular} & \begin{tabular}[c]{@{}c@{}}15\\ $mean\pm std$\end{tabular} & \begin{tabular}[c]{@{}c@{}}20\\ $mean\pm std$\end{tabular} & \begin{tabular}[c]{@{}c@{}}30\\ $mean\pm std$\end{tabular} & \begin{tabular}[c]{@{}c@{}}40\\ $mean\pm std$\end{tabular}         \\ \hline
Kinematic Model               & $0.968\pm 0.006$ & $0.968\pm 0.008$ & $0.959\pm 0.010$ & $0.921\pm 0.021$ & $0.871\pm 0.031$ \\
EDA Model                    & $0.697\pm 0.015$ & $0.625\pm 0.027$ & $0.593\pm 0.034$ & $0.502\pm 0.013$  & $0.480\pm 0.024$ \\
Kinematic-EDA Fusion Model          & $0.976\pm 0.005$ & $0.974\pm 0.009$ & $0.963\pm 0.010$ & $0.924\pm 0.026$ & $0.880\pm 0.015$ \\
EDA-enhanced Kinematic Model & $\textbf{0.978}\pm 0.003$ & $\textbf{0.976}\pm 0.005$ & $\textbf{0.968 }\pm 0.005$ & $\textbf{0.927}\pm 0.005$ & $\textbf{0.906}\pm 0.005$ \\ \bottomrule
\end{tabular}}}}
\end{table*}

\begin{table*}[!h]
\centering
\caption{Results of models with different time spans after downsampling}
\label{tab:best_result after downsampling}
\renewcommand\arraystretch{1.2}
\scalebox{1}{
\resizebox{\linewidth}{!}{
{
\begin{tabular}{lccccc}
\toprule
  Time span $T_s$  & \begin{tabular}[c]{@{}c@{}}10\\ $mean\pm std$\end{tabular} & \begin{tabular}[c]{@{}c@{}}15\\ $mean\pm std$\end{tabular} & \begin{tabular}[c]{@{}c@{}}20\\ $mean\pm std$\end{tabular} & \begin{tabular}[c]{@{}c@{}}30\\ $mean\pm std$\end{tabular} & \begin{tabular}[c]{@{}c@{}}40\\ $mean\pm std$\end{tabular}           \\ \hline
Kinematic Model               & $0.836\pm 0.032$ & $0.838\pm 0.012$ & $0.847\pm 0.042$ & $0.870\pm 0.012$ & $0.871\pm 0.031$ \\
EDA Model                     & $0.494\pm 0.019$ & $0.489\pm 0.039$ & $0.489\pm 0.035$ & $0.497\pm 0.031$ & $0.480\pm 0.024$ \\
Kinematic-EDA Fusion Model          & $0.845\pm 0.034$ & $0.843\pm 0.020$ & $0.869\pm 0.028$ & $0.884\pm 0.016$ & $0.880\pm 0.015$ \\
EDA-enhanced Kinematic Model & $\textbf{0.867}\pm 0.026$ & $\textbf{0.855}\pm 0.029$ & $\textbf{0.874}\pm 0.028$ & $\textbf{0.888}\pm 0.020$ & $\textbf{0.906}\pm 0.015$ \\ \bottomrule
\end{tabular}}}}
\end{table*}

\begin{table*}[!h]
\centering
\caption{Results of models with different exposure times. $n$ represents the number of segments removed}
\label{tab:best_result with different exposure time}
\renewcommand\arraystretch{1.2}
\scalebox{1}{
\resizebox{\linewidth}{!}{
\begin{tabular}{lccccc}
\toprule
      & \begin{tabular}[c]{@{}c@{}}n=1\\ $mean\pm std$\end{tabular} & \begin{tabular}[c]{@{}c@{}}n=2\\ $mean\pm std$\end{tabular} & \begin{tabular}[c]{@{}c@{}}n=3\\ $mean\pm std$\end{tabular} & \begin{tabular}[c]{@{}c@{}}n=4\\ $mean\pm std$\end{tabular} & \begin{tabular}[c]{@{}c@{}}n=5\\ $mean\pm std$\end{tabular} \\ \hline
Kinematic Model & $0.908\pm 0.029$  & $0.909\pm 0.025$  & $0.917\pm 0.011$      & $0.922\pm 0.030$   & $0.925\pm 0.031$                                      \\
EDA Model       & $0.525\pm 0.034$  & $0.505\pm 0.019$  & $0.532\pm 0.016$   & $0.510\pm 0.033$    & $0.518\pm 0.040$                                      \\ \bottomrule
\end{tabular}}
}
\end{table*}

Results are presented in the following subsections. After model setup, Optuna was used for parameter optimization, and an example of fine-tuning results with a time span of 30s before downsampling is shown in our supplemental material~\footnote{Available on OSF at \href{https://osf.io/9tucz/?view_only=fcf2188c009446cca87aeac085b411d9}{osf.io}, released under a CC BY 4.0 license. }. Firstly, the impact of different time spans on the model was investigated in \autoref{Performance on different time span}. Next, the performance of different models was analyzed and compared in \autoref{Model Comparison}. The proposed EDA-enhanced Kinematic Model achieved the highest accuracy using only kinematic data. Lastly, the relationship between different exposure times and CS was studied in \autoref{Performance on different exposure time}.

\subsection{Performance on different time span} \label{Performance on different time span}

Time span $T_s$ refers to the length of recently recorded data, including physiological and kinematic data. We aim to study how the models can make predictions based on historical data from a certain period of time,. The time spans included 10s, 15s, 20s, 30s, and 40s, dividing the original 4-minute data into 24, 16, 12, 8, and 6 segments respectively. Retraining the model revealed that as the time span increased, the model's performance gradually decreased, as shown in \autoref{tab:best_result before downsampling}. For instance, in the EDA-enhanced Kinematic Model, the accuracy decreased from 0.978 when the time span was 10s to 0.906 when the time span was 40s.

\textbf{Results Analysis.} Surprisingly, the accuracy change contradicted our intuition that a longer time span should improve prediction performance by providing more information. We suspected the limited training sample size to be the bottleneck affecting model performance. At a 40s time span, the sample size is $6*157=942$, while at a 20s time span, it is $12*157=1884$. To ensure consistent training sample sizes, we randomly downsampled them to $6*157=942$ across all time spans. After retraining, results are shown in \autoref{tab:best_result after downsampling}. The Pearson correlation p-values between accuracy and time span were as follows: Kinematic Model ($r=.955, p=.011$), EDA Model ($r=-.520, p=.368$), Kinematic-EDA Fusion Model ($r=.875, p=.051$), EDA-enhanced Kinematic Model ($r=.933, p=.020$). Three p-values, close to or less than 0.05, suggested a correlation between accuracy and time span, aligning with our intuition. Comparing experiments before and after downsampling, we observed significantly reduced model performance under the same time span due to downsampling. For example, with a 20-second time span, the sample size decreased from $12*157=1884$ to 942, resulting in EDA-enhanced Kinematic model performance dropping from 0.968 to 0.874. Hence, training sample size became the model bottleneck.

\begin{figure}[h]
    \centering
    \includegraphics[width=0.6\columnwidth]{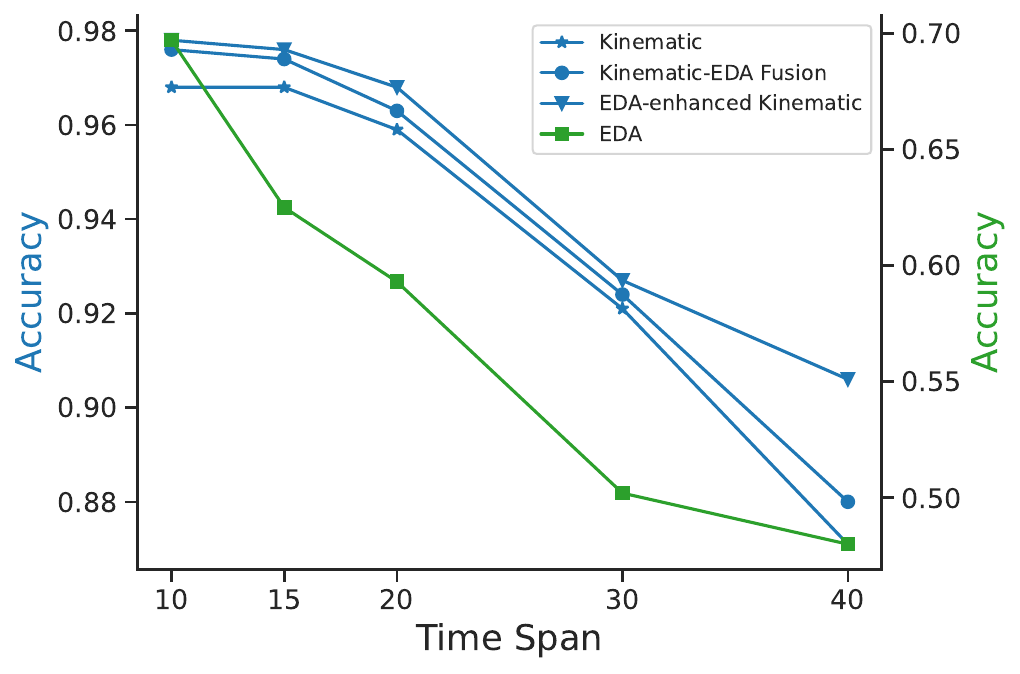}
    \caption{Performance on different time spans before downsampling.}
    \label{fig:Per_time span}
\end{figure}

In addition, we plotted the model performance curves before downsampling shown in \autoref{fig:Per_time span} and found that the three blue curves had a turning point at a time span of the 20s. Therefore, 20 is a value that balances both the training sample size and the amount of information carried.

\subsection{Model Comparison} \label{Model Comparison}

The accuracy of different models is presented in \autoref{tab:best_result before downsampling}. The results demonstrate that the proposed EDA-enhanced Kinematic Model can achieve the highest accuracy in predicting CS. Take the example when the time span is the 30s; specifically, the EDA model achieves an accuracy of 0.502 using only the EDA signals. When only using the kinematic data, the kinematic model achieves 0.921. The fusion model, which integrates physiological and kinematic data, has a slightly higher accuracy of 0.924 than the kinematic model. The proposed EDA-enhanced Kinematic Model achieves the best accuracy of 0.927, where kinematic and physiological embeddings were encoded using kinematic data. Hence, prediction is only based on kinematic data without utilizing physiological data.

\textbf{Results Analysis.} As shown in \autoref{tab:best_result before downsampling}, the EDA-enhanced Kinematic Model has the highest accuracy across the five time spans. After a $\chi^2$ test, we found a significant correlation ($p<.01$), indicating the effectiveness of the EDA-enhanced Kinematic Model in CS prediction. Fusion data outperforms single-modality models, highlighting the importance of leveraging multiple data modalities. Physiological and kinematic data contain valuable information for CS, enhancing integrated performance. Notably, the kinematic model outperforms the EDA model by 27\% when the time span is 10s before downsampling. The inferior EDA performance may be due to noise interference from motion artifacts, muscle activity, and environmental factors, impacting the accuracy and reliability of physiological signals. In contrast, kinematic data provides reliable information on body movement, less affected by noise, improving CS prediction.

\subsection{Performance on different exposure time} \label{Performance on different exposure time}

Exposure time refers to the duration of VR immersion. In \autoref{Problem Definition}, we assumed a constant CS level throughout the four-minute VR immersion. However, users may gradually experience CS during the immersion. To analyze the relationship between exposure time and CS, we truncated the four-minute experimental time, removed initial data, and retrained the Kinematic model and EDA model. As mentioned in \autoref{Performance on different time span}, using a 20s time span balances information and training data. We used it to study the exposure time-CS relationship. Truncating the first $n$ time segments (20s each) yields data from $n20$ seconds to 240 seconds. For instance, for $n=3$, we obtained data from 60s to 240s per sample, with a total size of $9*157=1413$. To ensure consistent training sample sizes, we randomly sampled to make them all $7*157=1099$. \autoref{tab:best_result with different exposure time} presents prediction results. Increasing $n$ improved the Kinematic model's performance. Accuracy increased from 0.908 (when $n=1$) to 0.925 (when $n=5$).

\textbf{Results Analysis.} From \autoref{tab:best_result with different exposure time}, interestingly, after conducting Pearson correlation tests between $n$ and accuracy, we found that the Kinematic model showed a significant correlation ($r=.978, p<.01$) and its performance gradually improved as $n$ increased. In contrast, the EDA model showed no significant correlation ($r=-.130, p=.834$) and its performance fluctuated as $n$ increased. High accuracy and correlation coefficient suggest that kinematic data maintains a high level of correlation with the CS throughout the entire four minutes of VR exposure. This correlation gradually increases as the exposure time lengthens.
In contrast, EDA signals showed a weak correlation with the CS, possibly due to signal noise. Overall, there is a certain correlation between exposure time and CS. Users are likely to experience a low CS level at the beginning of exposure, with the CS level gradually increasing as the exposure time lengthens. The experiment is designed as a simple case for navigation and further research is needed to reveal the relationship between exposure time and CS.

\section{Discussion}



In this section, we discuss the feasibility of our proposed prediction approaches. We will start by examining the prediction performance with EDA data, which is discussed in \autoref{Model Performance}. Then, in \autoref{benefits}, we demonstrate the benefits of our EDA-enhanced kinematic model, such as its reliance only on integrated sensors within HMDs and its ability to take into account physiological features. Finally, we discuss the limitations of our work in \autoref{limitations}.

\subsection{Prediction performance with EDA data} \label{Model Performance}

Our EDA feature extraction significantly improved prediction performance using EDA data. While some studies have used multimodal data, including EDA, few have focused solely on EDA for prediction. Compared to these studies, our EDA model achieved higher accuracy. For instance, Hadadi \textit{et al.} employed topological data analysis (TDA) techniques on physiological data, including EDA, TEM, and BVP, achieving an SVM classifier $F1$ score of 0.71 on binary classification~\cite{hadadi2022prediction}. Similarly, we used TDA to extract EDA features, which are included in our EDA numerical features. Our EDA-only model achieved a maximum accuracy of 0.697 and an $F1$ score of 0.690, which is impressive given the four-class classification task. Islam \textit{et al.} collected physiological data, including HR, EDA, and breathing rate, and calculated time series properties (e.g., max, min, avg, percentage change), achieving a high accuracy of 97.44\% for CS detection using a convolutional long short-term memory three-class classifier~\cite{islam2020automatic}. Their high accuracy can be attributed, in part, to their different data processing methods. They divided the data into 120-second time spans with a rolling window of 1 second, resulting in multiple overlapping time series data, which differs from our dataset.

\subsection{Benefits of EDA-enhanced Kinematic Model} \label{benefits}

\textbf{Usage of integrated sensors within HMDs.} 
Our motivation is to use fewer data dimensions collected from limited sensors in the headset to achieve higher accuracy. While many studies for CS prediction involving physiological signals often required external biosensors, our EDA-enhanced kinematic model only counted on integrated sensors within HMDs in a non-intrusive manner. Islam \textit{et al.} proposed a multimodal deep fusion approach for forecasting CS onset using the user's physiological, head-tracking, and eye-tracking data, which had to use eye-tracker and biosensors~\cite{islam2022towards}. Similarly, Qu \textit{et al.} utilized real-time physiological signals such as EDA and electrocardiogram and position and bone rotation data of users' virtual avatars to train an LSTM Attention neural network model~\cite{qu2022bio}. Although these models demonstrated remarkable prediction performance, the reliance on external sensors limited user freedom and practical usage. In contrast, our model only needed kinematic data from integrated sensors in HMDs, making it easier to be deployed in existing HMDs.

\textbf{Reflect individual difference.} 
Despite utilizing only the integrated sensors within HMDs like other studies, our EDA-enhanced kinematic model learned the physiological embedding from kinematic data to reflect individual differences. This gave our model a more comprehensive range of perspectives, improving its performance to predict CS. Islam \textit{et al.} proposed a deep fusion method that uses eye-tracking and head-tracking data to predict CS with an accuracy of 87.77\% and an RMSE of 0.51~\cite{islam2021cybersickness}. In addition, Du \textit{et al.} extracted saliency, optical flow, and disparity features from videos to determine the factors that contribute to simulator sickness. Their 3D CNN model showed improved performance in terms of Root Mean Square Error (RMSE) and Pearson Linear Correlation Coefficient.~\cite{du2021learning}. In contrast, our model specifically learns individual differences only from kinematic data, setting it apart from the above studies.

\textbf{Real-time Performance and applicable scenario.} 
Our EDA-enhanced Kinematic model stands out by demonstrating remarkable real-time performance, achieving an impressive accuracy of 97.8\% using only 10 seconds of historical kinematic data, as shown in Table \ref{tab:best_result before downsampling}. In comparison, Islam \textit{et al.} proposed a model that can forecast cybersickness onset 60 seconds in advance~\cite{islam2022towards}. Jin \textit{et al.} presented a machine learning approach to predict cybersickness level with the last 30-second data~\cite{jin2018automatic}.

Once the user wears the VR headset, our EDA-enhanced Kinematic model utilizes kinematic data collected from the headset to predict the occurrence of cybersickness. This prediction is essential as it enables the implementation of appropriate prevention strategies. By identifying whether a person is experiencing cybersickness, interventions such as reducing the field of view~\cite{fernandes2016combating, zielasko2018dynamic} or adjusting navigation speed~\cite{argelaguet2016giant} can be promptly employed.

\subsection{Limitations} \label{limitations}

There are three main limitations in our study.

\textbf{Data Collection and Processing.} 
For data collection, we note that data from integrated sensors in HMDs are not limited to motion and head-trackng but also include video and eye-tracking data, providing opportunities for further research in a multimodal manner~\cite{du2021learning, jin2018automatic}. We also find that our EDA model did not perform well during the cross-validation compared with the kinematic model, as the physiological signal is subject to noise and artifacts. Compared to motion and head-tracking data, EDA data contains more noisy information; thus, high-quality EDA data for training are encouraged.

In terms of data processing, although our EDA-enhanced kinematic model achieved the best performance by using head-tracking and motion data, the accuracy of our model was limited by the ground-truth construction, which assumes that the SSQ score collected after immersion denotes the CS level during the entire process. Therefore, we suggest that future studies use Fast Motion Scale (FMS) to collect more granular subjective measures of CS during immersion. 

\textbf{Model Architecture.} In comparison to previous studies~\cite{kim2019deep, islam2022towards, islam2021cybersickness}, our proposed approach achieves significant performance improvement using only LSTM and Dense layers. However, we notice that more complex networks, such as AttnLSTM and DeepTCN, have been employed by other researchers to classify the severity of CS. Therefore, we suggest further exploring such models to improve the performance of CS prediction. 

\textbf{Generality.} According to the sensory conflict theory, cybersickness is aroused by conflicting senses, which are aroused from anything that can stimulate visual-vestibular mismatch. So to enhance generality in our experiments, we enabled users to set their speed freely, thus providing different level of conflict and a way to generalize the findings beyond constant speed settings. However, we acknowledge that the navigation task employed in our study was a rudimentary simulation environment and that more complex simulations would be a necessary. Future research will contribute to a more comprehensive understanding of the effectiveness of our proposed method in predicting CS across different simulation environments.

\section{Conclusion}

In this work, we propose several prediction models for different data modalities. Among these models, the novel EDA-enhanced kinematic model achieves the highest accuracy of 97.8\% in CS prediction using only kinematic data, which learns physiological embedding with encoded physiological representation from the EDA model. To the best of our knowledge, this is the state-of-the-art performance for predicting CS using motion and head-tracking data. Note that our model can be practically applied to affordable consumer-level HMDs because it only relies on integrated sensors in HMDs. Our prediction model will allow future research to develop adaptive CS preventive measures to mitigate CS before the onset.


\bibliographystyle{abbrv-doi}

\bibliography{main}
\end{document}